%% file: ms.tex
\documentclass[apj,numberedappendix]{emulateapj_070502} 
\journalinfo{\rightline {To appear in The Astrophysical Journal}} 
\submitted{Received 2007 May 4; accepted 2007 August 31} 

\newcommand{\etal}{et al.}

\shorttitle{Diversity of SN I{\lowercase{a}}\ Rise Times}
\shortauthors{M.~Strovink}

\begin{document}

\title{Diversity of Decline-Rate-Corrected Type I{\lowercase{a}}\ Supernova Rise Times: \\ One Mode or Two?}  

\author{Mark Strovink} 

\affil{Physics Department and E.~O.~Lawrence Berkeley National Laboratory \\ 
       University of California, Berkeley, CA 94720}

\email{\tt strovink@lbl.gov}

\begin{abstract}
$B$-band light-curve rise times for eight unusually well-observed nearby Type Ia supernov{\ae} (SNe) are fitted by a newly developed template-building algorithm, using light-curve functions that are smooth, flexible, and free of potential bias from externally derived templates and other prior assumptions.  From the available literature, photometric {\it BVRI} data collected over many months, including the earliest points, are reconciled, combined, and fitted to a unique time of explosion for each SN.  On average, after they are corrected for light-curve decline rate, three SNe rise in $18.81 \pm 0.36$ days, while five SNe rise in $16.64 \pm 0.21$ days.  If all eight SNe are sampled from a single parent population (a hypothesis not favored by statistical tests), the rms intrinsic scatter of the decline-rate-corrected SN rise time is $0.96^{+0.52}_{-0.25}$ days -- a first measurement of this dispersion.  The corresponding global mean rise time is $17.44 \pm 0.39$ days, where the uncertainty is dominated by intrinsic variance.  This value is $\approx$2 days shorter than two published averages that nominally are twice as precise, though also based on small samples.  When comparing high-$z$ to low-$z$ SN luminosities for determining cosmological parameters, bias can be introduced by use of a light-curve template with an unrealistic rise time.  If the period over which light curves are sampled depends on $z$ in a manner typical of current search and measurement strategies, a two-day discrepancy in template rise time can bias the luminosity comparison by ${\approx}$0.03 magnitudes.     
\end{abstract}



\section{INTRODUCTION} \label{intro}

Within a few minutes of explosion, Type Ia supernov{\ae} release most of their energy, but due to self-absorption they reach peak luminosity only after 2-3 weeks.  During this period of ballistic expansion, while the photosphere grows in radius but shrinks in characteristic velocity as slower, heavier ejecta are revealed, basic properties of the explosion become evident.  Spectroscopic signatures of elements intermediate between carbon-oxygen fuel and iron-group ash (Filippenko 1997; Branch \etal\ 2006) reveal that burning is incomplete, with deflagration likely playing an early role \citep{Mazzali:2007}; nonvanishing polarization measures the progenitor's asphericity \citep{Wang:2007}.  After peak brightness, when iron features blanket the spectrum and polarizations wane, SNe become more homogeneous.  For SN science to progress, therefore, it is crucial to study the period of rising luminosity.  

In cosmological studies, SNe Ia are prized for their use as standardizable candles to trace the history of cosmic expansion (Riess \etal\ 1998, 2007; Perlmutter \etal\ 1999; for a review, see Perlmutter \& Schmidt 2003); in this context, periods of greater SN uniformity are of greater value.  Indeed, post-maximum luminosity indicators do yield low-dispersion Hubble diagrams (Wang \etal\ 2003; Wang, X. \etal\ 2005); post-maximum color measurements do solidify the corrections made for absorption by host-galactic dust (Lira 1995; Phillips \etal\ 1999; Jha \etal\ 2007).  Nevertheless, as ever more ambitious campaigns to chronicle the Universe's expansion history are planned ({\it e.g.}\ Aldering \etal\ 2002), the fundamental issue of high $z \to {\rm low} \; z$ evolution \citep{Howell:2007} keeps SN science in focus.  For example, more than one SN Ia progenitor or explosion mechanism might be at work, with progeny neither equally bright nor equally abundant at high {\it vs.}\ low redshift.  To secure such understanding, continued study of the rise-time period is essential.  

The subject of this report is a basic property of this period -- the light-curve rise time itself.  The $B$-band rise time is quite sensitive to the main-sequence mass of the white dwarf progenitor and to its carbon/oxygen ratio (see {\it e.g.}\ Dom{\'i}nguez \etal\ 2001).  It is less sensitive to the progenitor's metallicity.   

\citet{Pskovskii:1984} published the first rise times for classes of type I SNe.  In retrospect their range is reasonable, but, oddly, the reported correlation of rise time with decline rate was positive.  For individual SNe, the earliest rise-time measurements were made by \citet{Leibundgut:1991} (SN 1990N) and by \citet{Vacca:1996} (SN 1994D).  For a group of SNe, the earliest measurement of the average decline-rate-corrected rise time was reported by \citet{Groom:1998} and \citet{Goldhaber:1998}.  They used a model described {\it e.g.}\ by \citet{Arnett:1982}, in which the initial rise of $B$ flux with time is parabolic.  Within $\pm$ $\approx$2 days, these early determinations agree with current values.  Soon thereafter, in a definitive paper, \citet{Riess:1999b} (henceforth Rie99b) established the presently accepted average rise time to $B$ maximum of $19.5 \pm 0.2$ days.  This was accomplished by ferreting out early unfiltered photometry for ten nearby SNe and transforming it to standard passbands, to which the parabolic model was applied.  Also using this model, \citet{Conley:2006} (henceforth Con06) recently confirmed the Rie99b low-redshift average, in addition measuring an average rise time at high redshift of $19.10^{+0.18}_{-0.17}$(stat) $\pm 0.2$(syst) days.  Summarized later in this paper are additional measurements that quote larger uncertainties and/or apply to only 1-3 SNe; these best-fit rise times tend to be 1-3 days shorter.   

In this paper, using literature SNe, a first analysis is made of the intrinsic variation of SN rise times, {\sl after} correcting for their correlation with light-curve width.  Encountered are two curiosities: a remarkable distribution of rise-time variations and a surprising departure from the presently accepted average SN rise time.  Applied to this study is a newly developed light-curve-template-building algorithm that matches smooth, flexible functions to photometric data over many months including the earliest useful points.  Externally derived templates and other prior assumptions are avoided.  The method combines and reconciles the available {\it BVRI} photometry and fits it to a unique time of explosion for each SN.  Included in the analysis are three SNe with newly available early photometry.   

The bulk of this report is organized into five sections.  In \S\ref{defin} the rise time is defined, including its extrapolation to the time of explosion.  Described thereafter is the light-curve-template builder and its application to rise-time measurement.  In \S\ref{templ} the SN sample and its corrections are introduced.  Next a measure of SN light-curve width is chosen and its fitted values are discussed.  The fitted rise times then are presented and their uncertainties are analyzed.  In \S\ref{stati} the rise times are corrected for light-curve width, and their intrinsic variation is derived and characterized.  Particular attention is devoted to the possibility that the parent rise-time distribution is bimodal.  The discussion in \S\ref{discu} begins by addressing the issue of consistency with previous average rise-time measurements.  Thereafter the light-curve-width-corrected rise time is identified as a second SN Ia parameter, and its correlations with other variables of interest are surveyed.  Finally, \S\ref{impli} explores the implications for precision cosmology of both the intrinsic scatter in the rise time and the present inconsistencies in its measured average.

\section{RISE-TIME DEFINITION AND FITTING METHOD} \label{defin}

\subsection{Approach to Rise-Time Fitting} \label{appro}

Following \citet{Vacca:1996}, data including the earliest available observations are fitted to a single light curve in each band that everywhere satisfies stringent requirements on continuity, smoothness and analyticity.  Not pursued here is the piecewise method introduced by Rie99b, in which data collected earlier or later than a particular cutoff time $t_{\rm join}$, typically 10 days before the time $t_{B{\rm max}}$ of maximum $B$ flux, are fitted to dissimilar functions in the two regions.  In one variant of that method, at $t_{\rm join}$ the two fitted curves were not required to meet (Rie99b; Con06).  In another variant \citep{Aldering:2000}, the two curves did meet at $t_{\rm join}$, but the derivatives there were different.

Most of the analysis discussed by Rie99b, and all of the analysis by \citet{Aldering:2000} and Con06, is devoted to combined data samples from ensembles of SNe, many of which were not measured completely enough to provide well-defined individual {\it BVRI} light curves without imposing external templates.  Conversely, in order to measure the distribution of SN rise times, this analysis uses no templates and fits only individual SNe that do meet this standard.       

\subsection{Characterizing Light-Curve Width} \label{chara}

\citet{Goldhaber:1998}, Rie99b, and Con06 have established that SN Ia rise times are correlated positively with light-curve widths, fitting the latter using data collected no earlier than $\approx$10 days before $t_{B{\rm max}}$ (see also Riess \etal\ 1999a).  In broad use are three ``first parameters'' that characterize light-curve width: $\Delta m_{15}$ (Phillips 1993; Phillips \etal\ 1999); the {\sc mlcs} parameter $\Delta$ (Riess \etal\ 1995, 1996; Jha \etal\ 2007); and time-axis stretch $s_B$ (Perlmutter \etal\ 1997; Goldhaber, Groom \etal\ 2001); see also the parameters $t_{\pm 1/2}$ defined by \citet{Contardo:2000}.  For cosmology, light-curve width is used primarily to apply the empirical broader-brighter correction that relates the observed luminosity of a SN to its luminosity distance.  If that correction is based on both the rising and falling parts of the light curve, as is true for $\Delta$ and $s_B$, often its reliability is enhanced:  measuring the time of peak flux, as is required to determine $\Delta m_{15}$ without the aid of a template, is a delicate task (\S\ref{decli}).  

Nevertheless, to study the correlation between light-curve rise time and width, it is advantageous here to adopt a width parameter that, like $\Delta m_{15}$, is based only on the declining part of the curve.  Otherwise, part of the measured correlation would need to be attributed to the mutual redundancy of rise time and width.  Another advantage of $\Delta m_{15}$ is that it is a simple property of the light curve (the gain in $B$ magnitude between $t_{B{\rm max}}$ and 15 rest-frame days thereafter), which, for well-sampled SNe, can be determined without using a template (though usually it is not).  Conversely, for example, $s_B$ is measured by fitting the $B$ photometry to a single rigid template (Leibundgut 1988; Goldhaber, Groom \etal\ 2001).  When the photometry is exceedingly precise, the natural diversity of SN light-curve shapes can prevent template fits from achieving an acceptably probable $\chi^2$.  The best-fitted $\Delta$ or $s_B$ then may vary systematically, depending on when and how accurately the SN flux was sampled.  For this study, therefore, $\Delta m_{15}$ is a natural choice for measurement of SN light-curve width.

In this paper it is convenient to re{\"e}xpress the decline rate $\Delta m_{15}$ in terms of a tightly coupled quantity that has the same dimensions and order of magnitude as the rise time $t_{\rm r}$.  A natural choice is the functional inverse of $\Delta m_{15}$:  the fall time $t_{\rm f}$ is defined as the interval after $t_{B{\rm max}}$ (in rest-frame days) that is required for the $B$ magnitude to dim by 1.1 magnitudes ($1^{\rm m}\!\!.1$), the fiducial value of $\Delta m_{15}$ used by \citet{Phillips:1999} and many others.  Correspondingly, the fiducial fall time is 15 days.  The fall time is the same as the quantity $t_{+1/2}$ introduced by \citet{Contardo:2000}, except that it is referenced to a slightly later epoch (in their notation, $t_{\rm f}$ would be called $t_{+0.36}$).  Over the range of $\Delta m_{15}$ appropriate to SNe discussed in this paper, $t_{\rm f}$ is $>99.9\%$ correlated with $\Delta m_{15}$; the two variables are related by
\begin{equation}
t_{\rm f} \approx 15 - 8.42(\Delta m_{15}\!-\!1.1) 
  + 4.54(\Delta m_{15}\!-\!1.1)^2 
\, . \label{eqt_f}
\end{equation}
Also used here is the rise-time - fall-time difference $t_{\rm rf} \equiv t_{\rm r} - t_{\rm f}$.

Occasionally, for displaying and discussing the properties of $B$ and $V$ light curves, this paper uses a time dilation factor $s \equiv t_{\rm f}/15 \; {\rm days}$ that likewise is $>99.9\%$ correlated with $\Delta m_{15}$.  This factor is used only to define the dilated rest-frame phase $\tau$, 
\begin{equation}
\tau \equiv {{t - t_{B{\rm max}}} \over {s(1+z)}} 
\, , \label{eqtau}
\end{equation}
so that $B(\tau \!=\! 15 \; {\rm days}) - B(\tau \!=\! 0) \equiv 1^{\rm m}\!\!.1$.  Conveniently, when $B$ is plotted {\it vs.}\ $\tau$, all light curves exhibit the fiducial decline rate.

Finally, to make a rough estimate of the sensitivity of existing cosmological analyses to uncertainties in the rise time of the light-curve templates they have adopted, it is proposed (\S\ref{width}) that the stretch $s_B$ that is measured in these analyses is approximately proportional to a linear combination $t_{\rm r} + \gamma t_{\rm f}$ of rise and fall times.  That is, for an ensemble of SNe, a plot of $t_{\rm r}$ or $t_{\rm f}$ {\it vs.}\ $s_B$ would scatter about a straight line with finite slope.  This stands to reason: all three variables are amplified by the same factor if the time axis is dilated.  Because the declining part of the light curve is longer and may have smaller photometric errors than the rising part, one might expect $s_B$ to be more strongly correlated with $t_{\rm f}$ than with $t_{\rm r}$; correspondingly, $\gamma$ is assigned the value 2 in the examples given.

\subsection{Extrapolating to Time of Explosion} \label{extra}

Conventionally, the SN Ia rise time $t_{\rm r}$ is defined as the interval between the time of explosion $t_{\rm expl}$, when fluxes are assumed to vanish, and $t_{B{\rm max}}$.  When this definition is adopted, as it is here, measuring $t_{\rm r}$ requires extrapolating to $t_{\rm expl}$.  This is necessary because available template-quality filtered photometry for nearby SNe begins at most $\approx$3 magnitudes below peak.  If the initial time dependence of the $B$ flux is parabolic \citep{Arnett:1982}, for typical $B$ light curves $t_{\rm expl}$ occurs at least $\approx$3 days before the first datum.

When a particular function such as a parabola is chosen for this $>$3-day extrapolation in the $B$ band, the corresponding function for other bands can have exactly the same form only if the SN color does not evolve within the region of extrapolation.  Unfortunately, as is exhibited clearly in Fig.~7(b) of \citet{Pastorello:2007b}, this assumption is not fully supported by data available at slightly later epochs.  To explore this point, consider the rate of change $\nu_X \equiv {\rm d}(B\!-\!X) / {\rm d}\tau$ of $B\!-\!X$ color, where $X$ = $V$, $R$, or $I$, in the period $\tau < -11$ days.  In this analysis,  $\nu_X$ is parametrized by a constant, whose value is determined primarily by smooth extrapolation of the fit to $B\!-\!X$ data in the region $-11 < \tau < -5$ days.  Considering as diverse examples SN 1998aq, SN 1994D, SN 2002bo, and SN 2005cf, respectively, the best-fit $\nu_V$ is approximately $0^{\rm m}\!\!.000$, $-0^{\rm m}\!\!.011$, $-0^{\rm m}\!\!.045$, and $-0^{\rm m}\!\!.067$ dy$^{-1}$.  For the last three SNe, the departure from $\nu_V = 0$ is significant and nonnegligible compared to the maximum value ${\rm d}(B\!-\!V) / {\rm d}\tau \approx +0^{\rm m}\!\!.06$ dy$^{-1}$ typically reached during the post-$B_{\rm max}$ period $15 < \tau < 20$ days.  If the $B$-band flux rises parabolically,
\begin{equation}
f_B \propto (\tau-\tau_{\rm expl})^2, \label{eqfB}
\end{equation}
the flux in the $X$ band will exhibit a slightly slower rise,  
\begin{equation}
f_X \propto (\tau-\tau_{\rm expl})^2 \, 10^{0.4 \nu_X(\tau -\tau_{\rm expl})}, \label{eqfX}
\end{equation}
when $\nu_X$ is negative.

Using a parabola for the $B$-band extrapolation to $t_{\rm expl}$ is only one possible method for deriving $t_{\rm r}$ from the observed data.  It is a simple and conventional method, applied uniformly to each SN, that conforms to the earliest available data.  Should it be discovered that, at epochs too early to be well measured at present, the $B$ flux does deviate from parabolic time dependence, rise times that are corrected for that effect would shift systematically relative to those reported here.         

\subsection{The {\sc aquaa} Template Builder} \label{aquaa}

For constructing light-curve templates, a primary challenge is to devise functions that approach the smoothness of the parametrization used by \citet{Vacca:1996}, while attaining the flexibility required to follow the precise photometry in detail.  To achieve this flexibility, in earlier approaches cubic splines have been fitted to magnitudes (Hamuy \etal\ 1996; Prieto \etal\ 2006) or to fluxes \citep{Knop:2003}.  Because magnitudes soar near the time of explosion, the former type of fit has found use mainly after $t_{B{\rm max}} - 5$ days; because splines do not naturally mimic exponential decay, the latter type of fit has proved awkward in the region beginning $\approx$50 days after $t_{B{\rm max}}$, requiring an exponential to be spliced in.  As well, if the cubic spline knots are placed arbitrarily, unwanted ringing can arise.

Since August 2005 the template builder {\sc aquaa} ({\sc a}daptive {\sc qua}rtic {\sc a}lgorithm) has been under development, with the primary aim of producing full {\it (U)BVRI} light-curve templates representing individual SNe (Strovink, M.\ 2007, in preparation).   {\sc aquaa} is based on fits to quartic splines, which are smoother than cubic splines (third derivatives are continuous).  Further smoothing is gained by using ``fuzzy knots'' that allow gradual transitions between spline segments; the resulting curves are all-orders differentiable.  Knot positions are determined adaptively and objectively by the fit itself, subject only to minimum separation requirements that are SN-independent.

A quartic spline of this type is used to model the measured $B$-band flux.  In its earliest segment, a parabolic rise is imposed by requiring other coefficients to vanish.  Near-exponential decay in the region $\approx$$50 < \tau < 125$ days is modeled by using as the independent variable 
\begin{equation}
u = u_0 \tanh{\left((t - t_{B{\rm max}})/ u_0 \right)} \; , \label{equ0}
\end{equation}
where $u_0 \approx 125$ days, in place of the time $t$ itself.  In the spline's latest segment, the cubic and quartic coefficients are set to zero.

To fit the {\it VRI} bands,\footnote{For a few SNe with sufficient data, the $U \!-\! B$ color may similarly be modeled.  The $U$ band is not used here for rise-time determination.} similar quartic splines are used to model the $\bv$, $B\!-\!R$, and $B\!-\!I$ colors.  This choice is made because color curves have finite asymptotes; they vary less violently than magnitude curves; and, in the $I$ band, they have fewer extrema.  Implicitly it is assumed that colors evolve smoothly even near the (common) time of explosion, where magnitudes change violently; when early filtered data are scarce, this assumption enhances the fitted rise-time accuracy.  All coefficients for all splines are fitted simultaneously by minimizing a global $\chi^2$ that compares modeled to measured {\it BVRI} magnitudes and their quoted errors (assumed independent band-to-band).  Because the spline coefficients are overconstrained by the asymptotic requirements, the global fit requires the use of Lagrange undetermined multipliers.

Typically an average of 9 degrees of freedom (dof) per band are determined by the global fit.  In advance, scores of simpler fits are performed in order to zero in on the global optimum.  Currently {\sc aquaa} requires at least one day of human effort and CPU time to fit one SN.

\subsection{Applying {\sc aquaa} to Rise-Time Measurement} \label{apply}

\begin{figure*}
\epsscale{1.08}  
\plotone{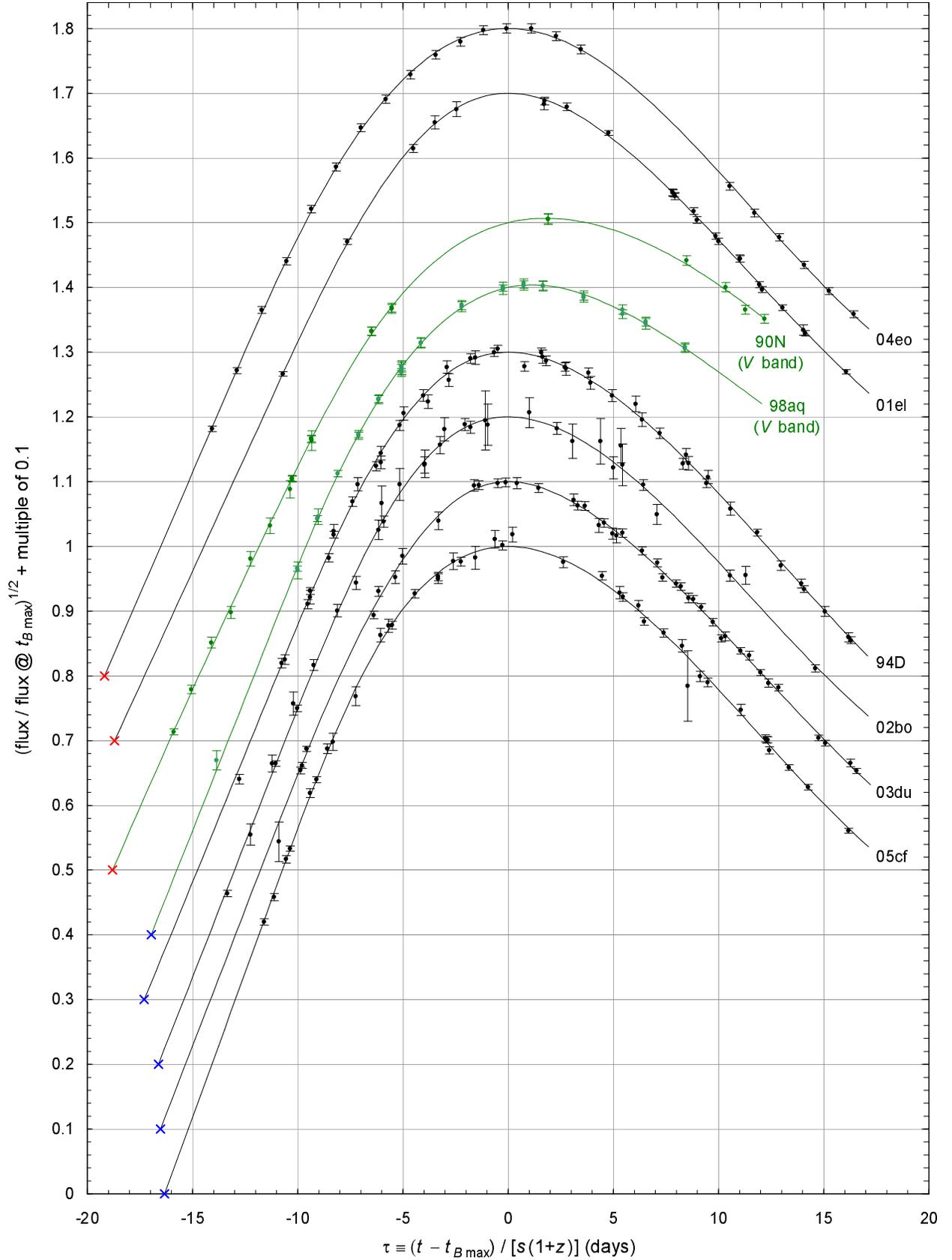}  

\caption{{Representative {\sc aquaa} light-curve fits.  The abscissa $\tau$ is the SN phase relative to $t_{B{\rm max}}$, dilated by a factor $s$ so that $\Delta m_{\tau=15} \equiv 1^{\rm m}\!\!.1$ in $B$ band.  The ordinate is the square root of the flux relative to its value at $\tau = 0$, offset by a multiple of 0.1.  $B$-band data are shown for SN 2005cf, SN 2003du, SN 2002bo, SN 1994D, SN 2001el, and SN 2004eo; $V$-band data are shown for SN 1998aq and SN 1990N.  The intersections of the fitted curves with zero flux, shown by ``$\times$'', are the best-fitted $s$-corrected times of explosion.}} \label{fig1}
\end{figure*}

Figure \ref{fig1} shows {\sc aquaa} fits to the eight literature SNe fitted for this rise-time study.  Although the rise time is the product of a global fit to all four {\it BVRI} bands over available data out to $\tau = 125$ days, Fig.~\ref{fig1} exhibits only $B$-band fits to SN 2005cf, SN 2003du, SN 2002bo, SN 1994D, SN 2001el, and SN 2004eo -- and only $V$ fits to SN 1998aq and SN 1990N -- over $\tau < 17$ days.  For the latter two SNe, the $V$ band is shown in order to include the unfiltered points transformed to that band by Rie99b; the fitted $V$-band maxima are reached 1.2 and 1.9 rest-frame days, respectively, after $t_{B{\rm max}}$.  On the ordinate, the square root of the flux (normalized to its $\tau=0$ value) is displayed so that a parabolic rise from explosion appears as a straight line.  For ease of comparison, all SNe are forced to have $\Delta m_{\tau=15} = 1^{\rm m}\!\!.1$ by using $\tau$ from equation (\ref{eqtau}) as the abscissa.  Therefore, by construction, all $B$ light curves agree at $\tau=0$ and $\tau=15$ days.  Nevertheless, close inspection reveals detailed differences in the shapes of the declining parts of the curves: without constraint from any template, {\sc aquaa} draws an independent smooth curve through each SN's data.

Another feature evident from Fig.~\ref{fig1} is that all fits are ``good'' ($\chi^2 \approx {\rm \# dof}$).  This also occurs in part by construction.  For each combination $i$ of telescope and band that contributes to the SN dataset (telescopes are grouped if many contribute), both the zero point and the photometry error may be perturbed.  To perturb the zero point, an offset $\delta_i$ and an offset tolerance $\epsilon_i$ are introduced; initially all $\delta_i = 0$ and all $\epsilon_i$ are equal.  The {\sc aquaa} fitted $\chi^2$ is augmented by $\sum_i (\delta_i/\epsilon_i)^2$ and the optimum $\{ \delta_i \}$ are determined as part of the global fit.  This process is repeated several times, with the $\{ \epsilon_i \}$ adjusted iteratively so that the offset contribution to $\chi^2$ is balanced among telescopes and bands, with a total of $\approx$ 1/\#dof.  When only two telescopes are involved, this procedure assigns their zero-point $|$offsets$|$ equally; but when many are involved, the iteration assigns tighter tolerances to telescopes that are in better mutual agreement.  Typically, telescopes with significant influence on the fitted light-curve shapes are assigned offset tolerances of $0^{\rm m}\!\!.02$ or less.  Rarely, when the fit strongly prefers it, a small color-term offset is combined with the zero-point offset.  To perturb the photometry error, an error floor (typically of order $0^{\rm m}\!\!.015$) may be imposed, and/or a constant error (typically of the same order) may be added in quadrature to the quoted uncertainties.  Again the procedure is iterated, assigning the largest error augmentations to the telescopes with greatest scatter (relative to quoted errors), while leaving unmodified the errors on the smoothest data.

As for the rising part of the light curves, on the scale of Fig.~\ref{fig1} the root fluxes in the $B$ band grow in a manner that is indistinguishable from linear over the first $\approx$40\% of their rise times.  Thereafter, everywhere maintaining a continuous third derivative, the curves gradually turn over in order to match the data around the peak.  Close inspection reveals, however, that the $V$-band root flux fitted to SN 1990N has a slower-than-linear rise.  As was discussed in \S{\ref{extra}}, this occurs because the initial rate of change $\nu_V$ of $\bv$ color, based on the best {\sc aquaa} fit to the nearest available SN 1990N data, is significantly less than zero.  As will be discussed in \S{\ref{error}}, uncertainties in the the {\sc aquaa} fitted output take into account the error associated with this extrapolation.  Appendix \ref{detai} supplies more detail on fits carried out for individual SNe.

The striking feature of Fig.~\ref{fig1} is that the fitted values of $\tau$ at which the explosions occurred, indicated by ``$\times$'', appear to divide into two groups separated by $\approx$2 days.  The diversity of decline-rate-corrected rise times would be apparent even if no fitted curves were displayed; to this topic the discussion now turns.

\section{RISE TIMES OF EIGHT TEMPLATE SNe} \label{templ}

\subsection{SN Sample Characteristics} \label{sampl}

At present, individual SNe with published photometry having sufficient cadence and accuracy to define independent {\it BVRI} templates are available only at low $z$ ($< 0.02$).  The eight nearby literature SNe studied here are listed in the first column of Table \ref{table1}.  Five of these (SN 1990N, SN 1994D, SN 1998aq, SN 2001el, and SN 2002bo) are among the eight nearby SNe whose rise times were analyzed by Con06.  The first three in this list are among the ten SNe studied by Rie99b; in particular, SN 1990N had a large influence on that paper's rise-time results.  The analysis reported here does not use the four SNe studied by Rie99b but not by Con06, nor does it use three of the SNe studied in both papers.  For six of these seven SNe, the quality of the published photometry does not allow {\sc aquaa} to define {\it BVRI} templates independently for each SN.  For the seventh, SN 1998bu, the only published datum earlier than $\approx$4 days before $t_{B{\rm max}}$ is a single unfiltered point with a sizable uncertainty ($\approx 0^{\rm m}\!\!.15$).

To the five SNe in common with Con06, this analysis adds three that are more recently measured:  SN 2003du (Leonard \etal\ 2005; Anupama \etal\ 2005; Stanishev \etal\ 2007; Li, W.D. \etal\ 2007, in preparation); SN 2004eo (Hamuy \etal\ 2006; Pastorello \etal\ 2007a); and SN 2005cf \citep{Pastorello:2007b}.  These recent SNe are splendidly observed and contribute greatly to this report.

Excluding SN 2004eo, this sample has an average heliocentric redshift  ${\langle z \rangle} = 0.004$ with an rms $\sigma_z = 0.002$; such a low $\sigma_z$ introduces a negligible rise-time dispersion.  Following usual practice (see, for example, \citet{Prieto:2006}), SNe with $z < 0.01$ were not $K$-corrected; SN 2004eo, with $z = 0.016$, was $K$-corrected.  To first order, to compensate for dust absorption $A_X$ in band $X$, dereddening ($R$-) corrections add a constant offset $-A_X$, which has no effect on this analysis.  To second order, $R$-corrections perturb the light-curve shape.  This occurs mainly where colors are changing rapidly, affecting the $B$ light curve's falling part more than its rising part.  SN 2002bo and SN 2001el were $R$-corrected using the $A(\lambda)$ of \citet{Cardelli:1989}.  The other six SNe, having $\langle (\bv)(t_{B{\rm max}}) \rangle = -0^{\rm m}\!\!.01$ with an rms of $0^{\rm m}\!\!.08$, on average are not too far from $(\bv)(t_{B{\rm max}}) \approx -0^{\rm m}\!\!.05$, as is typical of fiducial SNe Ia, and so were not $R$-corrected.  The $B$ and $V$ data from Fig.~5 of \citet{Hamuy:2006} that contribute to the {\sc aquaa} fit of SN 2004eo were ($S$-) corrected to the standard filters of \citet{Bessell:1990} using the procedure of \citet{Stritzinger:2002}. 

In summary, three of the eight SNe received a correction:  SN 2001el ($R$); SN 2002bo ($R$); and SN 2004eo ($KS$).  The combined ($KRS$-) corrections were calculated using v1.2 of the spectral template series introduced by \citet{Nugent:2002}.\footnote{\url{\tt http://supernova.lbl.gov/\~{}nugent}}  Iteratively these templates were dilated and warped so that, after $KRS$-corrections, the synthesized $\bv$ color evolution matched the observed evolution.  Within the range of Fig.~\ref{fig1}, the largest $|KRS$-corrections$|$ to $B \!-\! B_{\rm max}$ for SN 2001el, SN 2002bo, and SN 2004eo, respectively, were small: $-0^{\rm m}\!\!.044$, $-0^{\rm m}\!\!.064$, and (owing to a cancellation) $+0^{\rm m}\!\!.009$.

All SNe that ever have been fitted by {\sc aquaa} are included in the rise-time sample, except for SN 1995D, SN 1996X, and SN 2002fk.  The first two are not useful for rise-time analysis due to lack of $B$-band data for $\tau < -4$ days.  For the last, which is similar in rise and fall time to two SNe in the rise-time sample, the photometry is not yet published (Li, W.D.\ \etal\ 2007, in preparation).  Therefore the sample is unbiased by foreknowledge of fitted rise-time values.  SNe belonging to the subsets ``1991T-like'' and ``1991bg-like'' (Filippenko \etal\ 1992a,b) that are equally well suited for rise time analysis were not identified and therefore not added to the sample.  Excluded were more peculiar SNe such as SN 2000cx, which exhibited unusual color evolution and spectral features \citep{Li:2001} along with an obviously short rise time.  The result is a standard SN Ia set similar to one that might be used in a precise cosmological study.  Its members are unusually well observed, possessing normal characteristics and requiring very little correction.    
      
\subsection{Fitted Decline Rates and Fall Times} \label{decli}

Care must be exercised in fitting the decline rate $\Delta m_{15}$: because the time derivative of the flux at day 15 is large, a small bias in the definition of $t_{B{\rm max}}$ can propagate into a large bias in $\Delta m_{15}$.  The {\sc aquaa} fitted decline rates appear in column 3 of Table \ref{table1}.  Whenever $\Delta m_{15}$ values and uncertainties are quoted in the photometry references tabulated there, they are found to be in acceptable agreement with the {\sc aquaa} fitted values.  For both SN 1990N and SN 1994D, similar concordance is found with values quoted both by \citet{Phillips:1999} and by Rie99b.

In Fig.~\ref{fig2} the filled squares show the {\sc aquaa} fitted fall times $t_{\rm f}$ {\it vs.}\ $\Delta m_{15}$ for this sample (with SN 1995D and SN 1996X added).  The exhibited best-fit quadratic curve through these points is given by equation \ref{eqt_f} (\S\ref{chara}).  When the ordinate is transformed so that this curve becomes a straight line, the filled squares in the transformed Fig.~\ref{fig2} fully correlate the ordinate with the abscissa (Pearson $R > 0.999$).

\subsection{Fitted Rise Times} \label{riset}

The {\sc aquaa} fitted rise times appear in column 4 of Table \ref{table1} and as filled circles in Fig.~\ref{fig2}.  In addition, plotted as an open triangle is the average fitted rise time for ten SNe corrected to unit stretch by Rie99b; shown as an open square is a similar average for the eight low-$z$ SNe studied by Con06.  These two points are plotted at the fiducial $\Delta m_{15}$ values that corresponded to unit stretch for each analysis (Rie99b; Conley, A., private communication).  Both statistical and systematic uncertainties are included in all error bars (see \S{\ref{error}}).

As is discussed in Appendix \ref{compa}, for five SNe the rise times in Table \ref{table1} may be compared to published determinations that used the piecewise method of Rie99b (see \S{\ref{appro}}).  Compared to {\sc aquaa} values, the piecewise method yields rise times that, on average, are longer by $1.28 \pm 0.33$ days.  Of this systematic difference, $0.44 \pm 0.15$ days are due to the later $t_{B{\rm max}}$ that was used by the published rise-time determinations.  The extreme instance of this ($t_{B{\rm max}}$ later by 0.9 days for SN 1990N) may be ascribed to the use by Rie99b of only a second-order polynomial to fit the $B$ peak.  If a higher-order polynomial is used, the empirical fact that $B$ light curves rise faster than they fall causes the fitted peak to become slightly asymmetric, yielding a slightly earlier $t_{B{\rm max}}$.  The balance of the systematically longer rise time obtained by the piecewise method occurs because that method assumes that $\bv$ color does not vary within the early piece, and because it does not require the two pieces to join.  This point is elaborated in Appendix \ref{compa}.

As foreshadowed by Fig.~\ref{fig1}, the striking conclusion from the {\sc aquaa} fitted rise times and decline rates plotted in Fig.~\ref{fig2} is that SN Ia rise times diverge significantly, in a way for which one cannot compensate by applying a purely decline-rate-dependent correction. 

\subsection{Error Analysis} \label{error}

A distinct advantage of using a single global fit to all the relevant ({\it BVRI}) data to determine a parameter such as the rise time is the simplicity of the (frequentist) approach used here to propagate photometric errors into the fitted parameter's uncertainty.  In the global fit, when the parameter in question is varied and the change in likelihood is recorded, all other ($\approx 4 \times 9$) free parameters are allowed also to vary for best compatibility with that excursion.  (One such free parameter, defined in \S{\ref{extra}}, is $\nu_V$, the coefficient of the linear term in the earliest segment of the quartic spline representing $\bv$ color.)  Then the curvature at maximum of the one-dimensional log likelihood function yields the photometric error on the parameter in question.  In this approach, as for all analysis in this report, no Bayesian priors are employed.  (If, alternatively, an integral over a multidimensional likelihood surface had been performed, implicitly a Bayesian prior would have needed to be applied.)

\input{tab1.tex}

As is often the case, here the most challenging aspect of error analysis is not the propagation of random errors, but rather the estimation of systematic errors.  In the development of {\sc aquaa}, many algorithmic choices were made.  For example, a variety of schemes for requiring a minimum degree of light-curve smoothness were explored.  Several of these schemes are plausible and defensible; nominally, choosing among them should have a neutral effect upon any fitted parameter.  Eventually a particular scheme was adopted.  However, despite their nominal neutrality, such algorithmic choices had discernible impacts on some fitted values.  Fortunately, over the long {\sc aquaa} development period, these fitted values were recorded regularly, for example since April 2006 for SN 1994D and SN 2001el.  For most SNe, these records make it possible, for example, to estimate a {\sl systematic} rms $t_{B{\rm max}}$ and rms $\Delta m_{15}$.  Conservatively, for a particular parameter, each SN is assigned a systematic error equal to the median systematic rms for that parameter, or its own, whichever is greater.

Particular attention has been devoted to systematic error on the rise time $t_{\rm r}$.  First, it should be emphasized that any systematic uncertainty in $t_{B{\rm max}}$ propagates directly into a systematic error in $t_{\rm r}$ (and in fall time $t_{\rm f}$).  If $t_{\rm f}$ is subtracted from $t_{\rm r}$, as will be discussed in \S\ref{corre}, this source of error doubles.  As a further check, for all but SN 2001el and SN2003du, the extent and quality of flux measurements at $\tau < -10$ days permit a meaningful parabolic fit to those data alone.  This cross-check was carried out in the $B$ filter only, except for SN 1990N and SN 1998aq, for which the $V$-filter data that largely control their times of explosion $t_{\rm expl}$ were included.  From these simple fits, $t_{\rm expl}$  on average differed by only +0.01 days from those of {\sc aquaa}, with an rms difference of 0.33 days.  Conservatively, for these SNe, the full difference in $t_{\rm expl}$ between the simple and {\sc aquaa} fits was added in quadrature to the other rise-time errors; for the remaining SNe, the median of those $|$differences$|$ was used.  Finally, the possibility of Malmquist bias in the measurement of $t_{\rm r}$ was discussed by Rie99b.  As will be seen, the analysis here is focused mainly on the $B$ light curve's symmetry, which to leading order is luminosity-independent.

\begin{figure}
\epsscale{1.17}  
\plotone{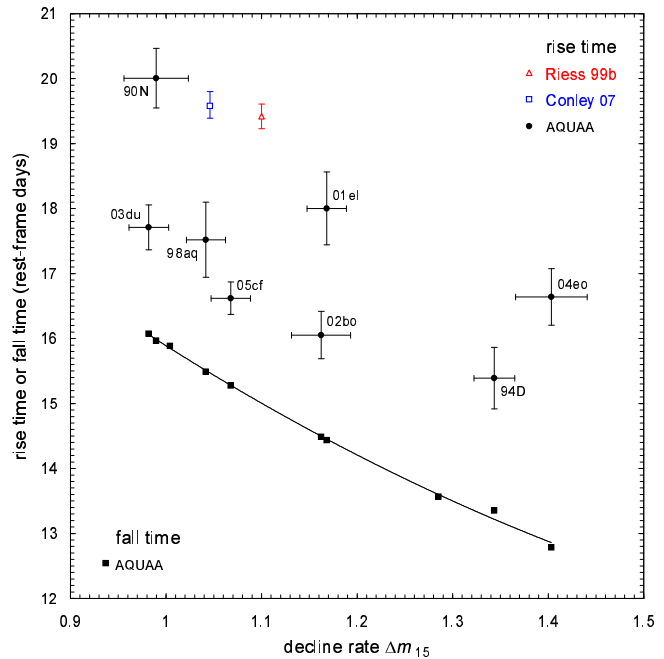}  
\caption{Fitted rise time or fall time (defined in the text) {\it vs.}~decline rate $\Delta m_{15}$.  Filled circles show the {\sc aquaa} fitted rise times for individual SNe.  For each SN, both statistical and systematic uncertainties are included in its horizontal and vertical error, which are positively correlated.  Filled squares show the {\sc aquaa} fitted fall times for the same SNe (plus two others with scant pre-maximum data).  The quadratic curve is drawn to guide the eye.  Open symbols show the results of previous rise-time fits by Rie99b (triangle) and Con06 (square) to ensembles of stretch-corrected low-$z$ SNe.} \label{fig2}
\end{figure}

In summary, the parameter errors listed in Table \ref{table1} and displayed in Figs.~\ref{fig2} and \ref{fig3} are the quadrature sum of ({\it i}) straightforwardly propagated photometric random errors; ({\it ii}) systematic errors estimated from records of {\sc aquaa} algorithm evolution; and, for rise times, ({\it iii}) differences in $t_{\rm expl}$ between {\sc aquaa} and simple parabolic fits to early data.   


\section{STATISTICAL PROPERTIES OF THE RISE-TIME SAMPLE} \label{stati}

\subsection{Correcting the Rise Time for Decline Rate} \label{corre}

As noted in \S{\ref{chara}}, \citet{Goldhaber:1998}, Rie99b, and Con06 independently found that SN Ia rise times are correlated positively with light-curve widths.  The anticorrelation of rise time with decline rate that is apparent in Fig.~\ref{fig2} supports these findings (although this support would lose statistical significance if SN 1990N were dropped as an outlier).  In the analysis that follows (which retains SN 1990N), the average rise time is referenced to a fiducial decline rate, and the portion of the rise-time variation that is independent of decline-rate variation is studied.  For these purposes, the functional dependence of rise time on decline rate needs to be characterized.  However, it is nontrivial to fit even a simple function to data whose absciss\ae\ and ordinates are not highly correlated, with errors that are nonnegligible fractions of their ranges (see, for example, the discussion by \citet{Wang:2006} of the fit to the data in their Fig.~6).

The approach taken here is to apply only a simple {\it ad hoc} correction for this anticorrelation -- a choice that is not necessarily optimal, but that does, within the statistical error, obviate the need for further correction.  The form chosen for this correction is motivated by the impression that, in Fig.~\ref{fig2}, the rise time and fall time exhibit a similar decline with increasing $\Delta m_{15}$.  Therefore, to correct the rise time for decline rate, the fall time is simply subtracted from it.  This difference $t_{\rm rf} \equiv t_{\rm r}-t_{\rm f}$ is exhibited in the fifth column of Table \ref{table1}, and it is plotted {\it vs.}~$\Delta m_{15}$ in Fig.~\ref{fig3}(a)\ and (b).  As seen in Fig.~\ref{fig3}(b), the points there do not substantially correlate $t_{\rm rf}$ with $\Delta m_{15}$ (Pearson $R = 0.25^{+0.36}_{-0.44}$).  Because of this feature, the statistical properties of $t_{\rm rf}$ can be studied without reference to a particular decline rate: the analysis discussed below assigns to $t_{\rm rf}$ a $\Delta m_{15}$-independent mean and dispersion.     

\subsection{Intrinsic Variation} \label{intrin}

As presented in Table \ref{table1} and shown by the error bars in Fig.~\ref{fig3}(a) and by the {\sl inner} error bars in Fig.~\ref{fig3}(b), the total error on $t_{\rm rf}$ is substantially larger than on $t_{\rm r}$.  As discussed in \S\ref{error}, this arises in part from the doubled effect of the uncertainty in $t_{B{\rm max}}$ upon $t_{\rm rf}$, compared to its effect upon $t_{\rm r}$.  Notwithstanding the larger error, an attempt to assign a unique best-fitted $t_{\rm rf}$ using these errors would yield an unacceptably high $\chi^2$ (29.4/7 dof, confidence level $1.2 \times 10^{-4}$).  Therefore, beyond the total uncertainty $\sigma^{\rm meas}_{\rm rf}$ with which it is measured, $t_{\rm rf}$ demonstrates a nonvanishing intrinsic variation $\sigma^{\rm int}_{\rm rf}$.  The sum in quadrature of $\sigma^{\rm int}_{\rm rf}$ and  $\sigma^{\rm meas}_{\rm rf}$ is shown by the outer error bars in Fig.~\ref{fig3}(b), which are dominated by $\sigma^{\rm int}_{\rm rf}$.  The quoted value 
\begin{equation}
\sigma^{\rm int}_{\rm rf} = 0.96^{+0.52}_{-0.25} \; {\rm days} 
\label{eqint}
\end{equation}
is obtained by varying $\sigma^{\rm int}_{\rm rf}$ to yield confidence levels of 0.5, 0.84, and 0.16, respectively, for fits to unique values of $t_{\rm rf}$.         
\subsection{Sample Average} \label{avera}

\begin{figure*}
\epsscale{0.85}  
\plotone{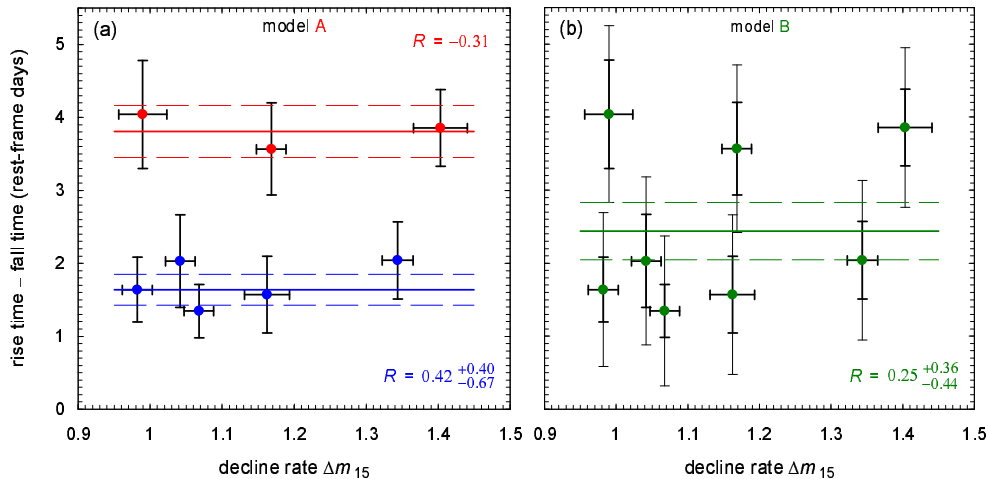}  
\caption{(a)\ Difference between the rise time and fall time shown in Fig.~\ref{fig2} {\it vs.}~decline rate $\Delta m_{15}$.  The error bars represent the combined statistical and systematic uncertainties.  Horizontal lines depict the ordinates best fitted to the slowest-rising three SNe (top) and to the fastest-rising five SNe (bottom) with their one-standard-deviation uncertainties.  The exhibited Pearson coefficients $R$ show that the abscissa and ordinate are not significantly correlated for either set of points.  (b)\ Same as (a) except that a single ordinate is fit to all eight SNe using the outer error bars, which include a 0.96-day intrinsic rise-time variation added in quadrature.}\label{fig3}
\end{figure*}

Using this central value of $\sigma^{\rm int}_{\rm rf}$, the best-fitted average difference of rise time and fall time is
\begin{equation}
\langle t_{\rm rf} \rangle = 2.44 \pm 0.39 \; {\rm days,} 
\label{eqavg}
\end{equation}
where the uncertainty is dominated by the intrinsic variance arising from $\sigma^{\rm int}_{\rm rf}$.  This value and its uncertainty are illustrated by the horizontal lines in Fig.~\ref{fig3}(b).  Correspondingly, a hypothetical large sample of SNe Ia exhibiting the fiducial fall time of 15 days would have a best-fitted average rise time
\begin{equation}
\langle t^{\rm fid}_{\rm r} \rangle = 17.44 \pm 0.39 \; {\rm days.} 
\label{eqfid}
\end{equation}


\begin{figure}
\epsscale{0.95}
\plotone{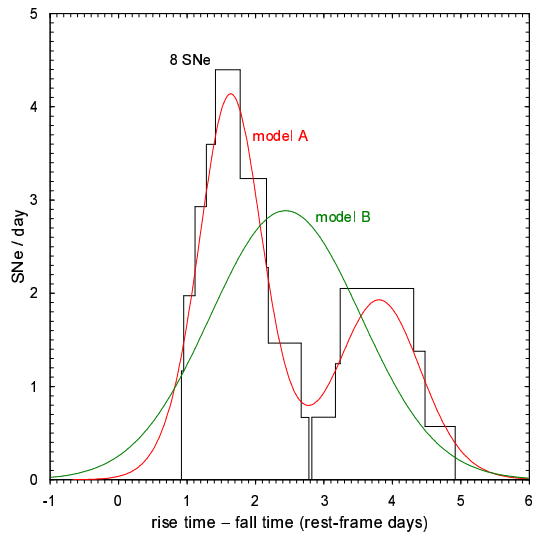}  
\caption{Unbinned histogram of differences $t_{\rm rf}$ between rise and fall times for 8 SNe.  Each SN is assigned a box of unit area and of width equal to its photometric FWHM resolution.  The unbinned histogram is the sum of the boxes.  Superimposed are expected summed probability density functions based on two different parent distributions convolved with the photometric resolutions.  The parent distributions are (model A) two $\delta$-functions in $t_{\rm rf}$, representing 5 fast- and 3 slow-rising SNe; and (model B) a single gaussian of width $\sigma^{\rm int}_{\rm rf}$.}\label{fig4}    
\end{figure}


\subsection{One Parent Population or Two?} \label{paren}

In Fig.~\ref{fig3}(a), separate values of $\langle t_{\rm rf} \rangle$ are fitted to the three slowest-rising and the five fastest-rising SNe, without augmenting the measurement errors to account for any intrinsic variance.  Obviously these fits are satisfactory.  The separate averages are
\begin{eqnarray}
{\rm slow}\!\!: \;\langle t_{\rm rf}\rangle &=& 3.81 \pm 0.36 \; {\rm days}; 
\nonumber \\ 
{\rm fast}\!\!: \;\langle t_{\rm rf}\rangle &=& 1.64 \pm 0.21 \; {\rm days}. 
\label{eqavg12}
\end{eqnarray}
Belaboring this point, Fig.~\ref{fig4} displays an unbinned histogram (defined in its caption) of the eight individual determinations of $t_{\rm rf}$.  Despite the meager statistics, it is difficult to view Fig.~\ref{fig3}(a) or Fig.~\ref{fig4} without considering the possibility that these SNe might be drawn from two different parent populations.

The present statistics do not support an exhaustive attempt to analyze all the circumstances that might account for the distributions in Figs.\ \ref{fig3} and \ref{fig4}.  For Monte Carlo simulation, only two simple models are considered here.  In model A, 3 SNe arise from one parent population, and 5 from a different population.  Each parent population has its own unique true value of $t_{\rm rf}$.  The differences between observed and true $t_{\rm rf}$ are drawn from normal distributions with rms values taken from the vertical error bars in Fig.~\ref{fig3}(a).  Conversely, in model B all 8 SNe arise from the same parent population, again having a unique true value of $t_{\rm rf}$.  Here the differences between observed and true $t_{\rm rf}$ are drawn from normal distributions using the {\sl outer} error bars in Fig.~\ref{fig3}(b).  In the simulated analysis of each 8-SN experiment, two free parameters are optimized for either model.  For model A, two values of $\langle t_{\rm rf} \rangle$ are fitted; for model B, one value of $\langle t_{\rm rf} \rangle$ is fitted, and $\sigma^{\rm int}_{\rm rf}$ is adjusted so that, after it is added in quadrature to the {\sl inner} error bars in Fig.~\ref{fig3}(b), $\chi^2$ = 1/dof for that fit. 

The present SN sample cannot discriminate definitively between models A and B.  Not only are the statistics inadequate, but these models and their statistical tests are contaminated by having been devised after the sample's statistical behavior had become evident.  Such {\it a posteriori} estimates of relative likelihood are notoriously unreliable.  Nevertheless, because of this second issue, it is still valuable here to discuss statistical tests of models A {\it vs.}\ B: in the future, when statistically independent samples of SNe for which individual rise times can be measured become available, the same models and tests described here may be reapplied without such contamination.

\begin{figure*}
\epsscale{0.85}  
\plotone{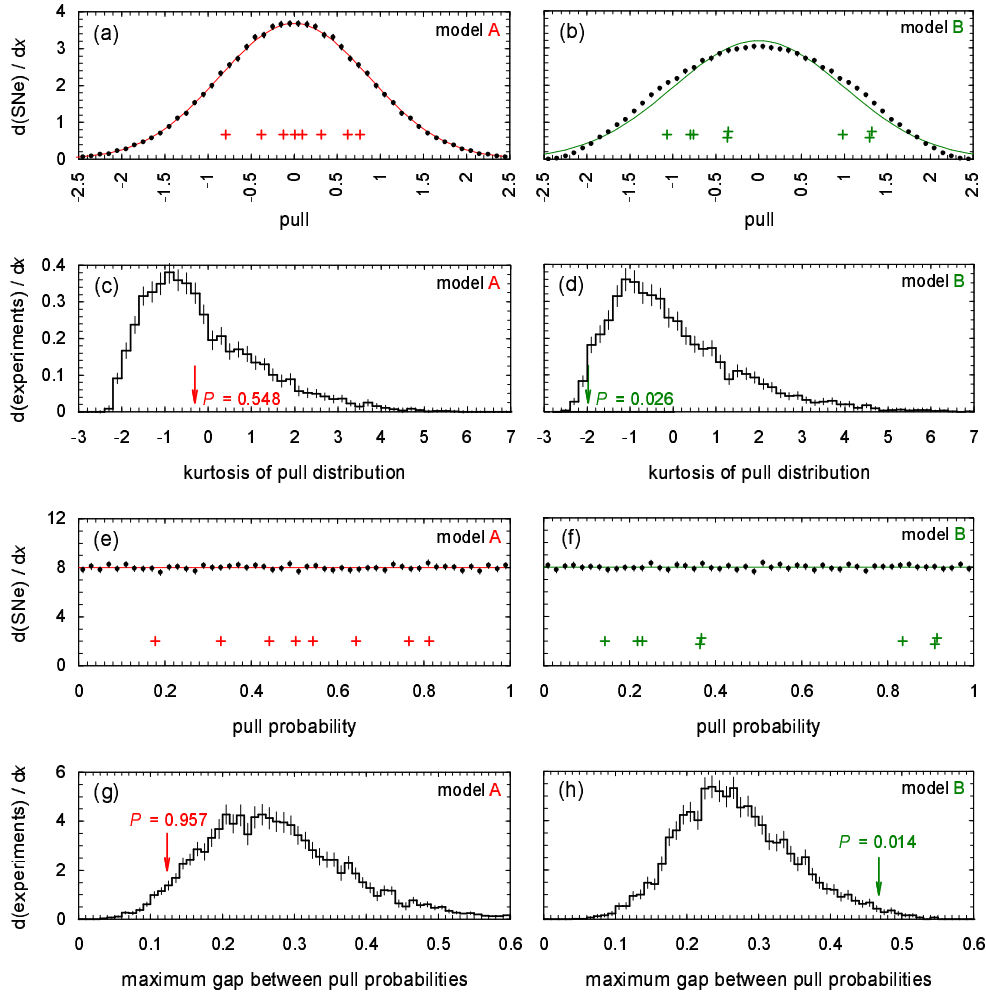}  
\caption{Comparison to gaussian expectation of the distribution of pulls (residuals normalized to their uncertainties) for the fits shown in Figs.~\ref{fig3}(a) (left column) and \ref{fig3}(b) (right column). (a)-(b)\ Distribution of pulls for eight actual SNe (crosses) and for fits to eight simulated SNe (points).  The gaussian curve has the same rms as the points.  (c)-(d)\ Kurtosis of the pull distribution for eight actual SNe (arrow) and for sets of eight simulated SNe (histogram).  (e)-(f)\ Same as (a)-(b)\ except that the pull is mapped to a function (``pull probability'') that is distributed uniformly on (0,1).  (g)-(h)\ Distribution of the maximum interval in pull probability between neighboring interior SNe, for eight actual SNe (arrow) and for sets of eight simulated SNe (histogram).}\label{fig5}
\end{figure*}

Figure \ref{fig5} illustrates two statistical tests of the relative likelihood that the actual eight-SN sample is described by model A {\it vs.}\ model B.  In this figure, the four plots on the left (right) refer to model A (B).  The crosses in Figs.~\ref{fig5}(a)-(b) show the distribution of the data in pull (the difference between the observed $t_{\rm rf}$ and its best-fit average, divided by the total uncertainty in $t_{\rm rf}$).  The points (with invisibly small uncertainties) show the Monte-Carlo-simulated pull distribution of SNe fitted in accord with one of the two models, and the curves are gaussians with the same rms width as the distribution of points.  Because $\chi^2$ is additive, the simulated pull distribution for model A is exactly gaussian.  For model B, however, the $\sigma^{\rm int}_{\rm rf}$ that is chosen for each simulated eight-SN experiment is affected by fluctuations in the values of $t_{\rm rf}$ that were generated for that experiment; therefore, as confirmed by Fig.~\ref{fig5}(b), the expected pull distribution is only approximately gaussian.


The usual estimator of population kurtosis,\footnote{\tt http://en.wikipedia.org/wiki/Kurtosis}
\begin{equation}
G_2 = {n-1 \over (n-2)(n-3)} \Bigl((n+1) {m_4 \over m_2^2} - 3 (n-1)\Bigr)
\; , 
\label{eqkurt}
\end{equation}
where $m_2$ is the sample variance and $m_4$ is the fourth sample moment about the mean, reaches a minimum for distributions consisting of two $\delta$-functions.  (A sample of $n = 8$ values divided 4:4 between two $\delta$-functions has $G_2 = -2.8$, while one divided 5:3 has $G_2 = -2.24$.)  Therefore $G_2$ is a natural estimator of a distribution's bimodality.  When analyzed in accord with model A, the eight actual SNe have $G_2 = -0.31$ -- a typical value -- while according to model B they have $G_2 = -1.98$.  These values appear as arrows in Figs.\ \ref{fig5}(c)-(d), to be compared to the simulated kurtosis distributions shown as histograms.  For model B, 26 in 1000 simulated experiments yield a kurtosis at least as small as the one obtained for the actual sample.    

The second statistical test requires mapping the pull into a ``pull probability'' variable that is distributed uniformly in the interval (0,1).  These uniform distributions are shown for simulated SNe (points) in Figs.\ \ref{fig5}(e)-(f); again the eight actual SNe appear as crosses.  For model B, between the fifth- and sixth-ranked actual SNe there is a large gap (0.47) in pull probability.  Such a gap is an estimator of bimodality that is different from (but not independent of) the kurtosis.  Figures \ref{fig5}(g)-(h) display histograms of the simulated distribution of maximum gap between adjacent pull probabilities, again with the actual values shown as arrows.  To construct these maximum gap distributions, gaps between SNe ranked 2 \& 3, 3 \& 4, 4 \& 5, 5 \& 6, and 6 \& 7 in pull probability were considered.   For model A, 957 in 1000 experiments would yield a larger gap; this may suggest that the systematic errors on $t_{\rm rf}$ were estimated too conservatively.  For model B, 14 in 1000 simulated experiments yield a gap in pull probability at least as large as the one obtained for the actual sample.      

\section{DISCUSSION} \label{discu}

Given the meager statistics, the topics chosen for the remaining discussion are relevant to any number of parent populations.

\subsection {Consistency of Measured Average Rise Times} \label{consi}

For SNe Ia with the fiducial decline rate, the {\sc aquaa} average rise time $\langle t^{\rm fid}_r \rangle = 17.44 \pm 0.39$ days is $\approx$2 days shorter than the stretch-corrected value $19.42 \pm 0.19$ days reported by Rie99b.  (Two other decline-rate-correction methods employed by those authors yielded essentially the same result.)  It is $\approx$2.1 days shorter than the stretch-corrected low-$z$ value $19.58^{+0.22}_{-0.19}$ days reported by Con06, and it is $\approx$1.7 days shorter than the high-$z$ rise time of $19.10^{+0.18}_{-0.17} \, ({\rm stat}) \pm 0.2 \, ({\rm sys})$ days reported there.  Using a well-observed nearby sample of similar size, and applying these observations efficiently to the measurement of $\langle t^{\rm fid}_r \rangle$, {\sc aquaa} nevertheless yields twice the low-$z$ rise-time uncertainty quoted in those papers.

About 2/3 of the inconsistency in central value is due to the tendency, discussed in \S\ref{riset} and in Appendix \ref{compa}, for the piecewise method of Rie99b to yield rise times longer than those obtained by {\sc aquaa}.  The balance of the central-value difference, as well as most of the difference in assigned uncertainty, is due to intrinsic variance.  Implicitly, it was assumed by Rie99b, and for low $z$ by Con06, that particular small samples, in which individual SNe such as SN 1990N carry heavy weight, accurately mirror the general SN Ia population; no error contributions from intrinsic variance were assigned.  As well, when a particular externally derived light-curve template was applied by those authors, implicitly it was assumed to represent accurately the average SN Ia behavior.  In fact, $B$ templates in wide use differ substantially in their asymmetry about $t_{B{\rm max}}$: $B(t_{B{\rm max}}+\Delta t) - B(t_{B{\rm max}}-\Delta t)$ can vary by up to one-half magnitude at $\Delta t \approx 10$ days.  Conversely, the analysis reported here applies no externally derived template, and its uncertainty on the average rise time is dominated by intrinsic variance.

Among additional determinations of SN Ia rise time, at low $z$ \citet{Aldering:2000} obtained the value $20.08 \pm 0.19 \, ({\rm stat})$ days.  Very recently, \citet{Garg:2007} reported a $V$-band rise time of $17.6 \pm 1.3 \, ({\rm stat}) \pm 1.1 \, ({\rm sys})$ days based on broadband observation of 3 SNe with redshifts $0.135 < z < 0.165$.  Since $V_{\rm max}$ usually occurs $\approx$1.5 days later than $t_{B{\rm max}}$, their central value would correspond to a $B$-band rise time of only $\approx$16 days.  At high $z$, \citet{Groom:1998} reported an average rise time of $17.6 \pm 0.4$ days based on a light curve to which 37 SNe contributed.  Applying a more complete error analysis to a large subset of those SNe, \citet{Aldering:2000} obtained $t_{\rm r} = 18.3 \pm 1.2 \, ({\rm stat}) \, ^{+3.6}_{-1.9} \, ({\rm sys})$ days, where the systematic errors are upper limits; see also \citep{Goldhaber:2001}.  When systematic as well as statistical errors are taken into account, these additional measurements are consistent with the rise time presented here.

\eject

\subsection {Yet Another Second Parameter} \label{anoth}

When spectroscopic features are taken into consideration \citep{Benetti:2005}, normal Type Ia SNe are no longer believed to be characterized broadly by a single ``first parameter'' such as $\Delta m_{15}$, $\Delta$, or stretch (see \S\ref{decli}).  More narrowly, a single parameter does not suffice even if attention is confined to the $B$ and $V$ light curves, which have relatively simple shapes.  For example, \citet{Jha:2007} obtain a true intrinsic color scatter in $(\bv)_{35}$, the $\bv$ color 35 days past $t_{B{\rm max}}$, equal to $\sigma_{(B\!-\!V){\rm int}} = 0^{\rm m}\!\!.049$.  Careful inspection of their Fig.~8(b) reveals that less than $10\%$ of $\sigma^2_{(B\!-\!V){\rm int}}$ can be ascribed to systematic dependence of $(\bv)_{35}$ on the {\sc mlcs}2k2 parameter $\Delta$.  Therefore $(\bv)_{35}$ is a second SN parameter.

Broadening the discussion to include the $R$ and $I$ bands, consider as a second example the ratio ${\cal R}^I_2 \equiv t_{\rm sec}/s_B$, where $t_{\rm sec}$ is the interval between $t_{B{\rm max}}$ and the second $I$-band peak, and $s_B$ is the $B$-band stretch.  Taking at face value the uncertainties in $t_{\rm sec}$ shown in Fig.~4 of \citet{Nobili:2005}, ${\cal R}^I_2$ exhibits substantial intrinsic scatter while remaining largely uncorrelated with $s_B$.  Therefore  ${\cal R}^I_2$ is another second parameter.  Many more second parameters may await identification.

Clearly the rise-time $-$ fall-time difference $t_{\rm rf}$ shown in Fig.~\ref{fig3}, exhibiting a $\approx$1-day intrinsic scatter and no significant correlation with $\Delta m_{15}$, by the same criteria is yet another second parameter.  

\subsection {Correlations with Other Variables} \label{other}

To understand better the factors controlling SN rise and fall time, it is interesting to measure the correlations of $t_{\rm r}$, $t_{\rm f}$, and $t_{\rm rf}$ with other SN variables.  For each parameter pair, this is done here by evaluating the Pearson correlation coefficient $R$, whose Fisher function $z' = \bigl(\ln{(1\!+\!R)}-\ln{(1\!-\!R)}\bigr)/2$ is distributed normally with variance $1/5$ for this eight-SN sample.

\begin{figure}
\epsscale{1.17}  
\plotone{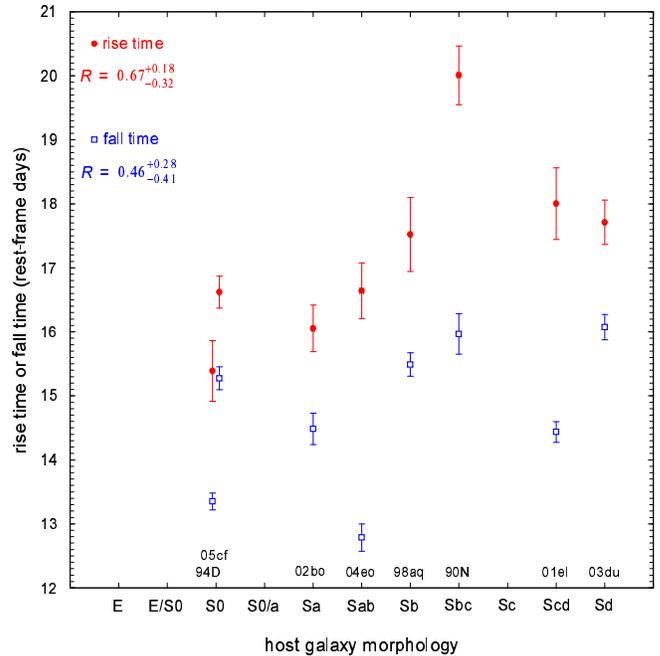}  
\caption{Fitted rise time (filled circles) or fall time (open squares), as in Fig.~\ref{fig2}, {\it vs.}\ host galaxy morphology.  Both statistical and systematic uncertainties are included in the error bars.  The positive Pearson correlation coefficient $R$ is more significant for rise time than for fall time.} \label{fig6}
\end{figure}

\citet{Mannucci:2007} describe two classes of Type Ia SN bimodality that are manifested in the properties of their host galaxies.  Used here is a simple measure of host-galaxy morphology: the integer $T$, ranging from $-3$ (E) to 7 (Sd),  as shown in Fig.~\ref{fig6}.  With respect to $T$, the rise time has a positive Pearson coefficient $R(t_{\rm r}, T) = 0.67^{+0.18}_{-0.32}$, while the fall-time correlation is not as significant, $R(t_{\rm f}, T) = 0.46^{+0.28}_{-0.41}$.  As expected, the correlation of stretch with $T$ is intermediate between that of $t_{\rm r}$ and $t_{\rm f}$.  \citet{Sullivan:2006} already have established at a higher statistical level the positive correlation of stretch with the rate of star formation.


Also examined here are the color indices $\bv$, $V\!-\!R$, $R\!-\!I$, and linear combinations thereof.  (Colors are measured at $t_{B{\rm max}}$ after correcting for Galactic dust using the map of \citet{Schlegel:1998}.)  Studied in addition is $\nu_V$, the initial rate of change of $\bv$ color (\S\ref{extra}).  For this small sample, no statistically significant correlations of these variables with $t_{\rm r}$, $t_{\rm f}$, or $t_{\rm rf}$ are evident.   

Not measured are correlations with absolute SN magnitude $M$: for only three of these SNe are distances well known (SN 2004eo is barely in the Hubble flow, and SN 1990N and SN 1998aq are {\it HST} Cepheid calibrated).  Despite their 2-day difference in $t_{\rm rf}$, SN 1990N and SN 1998aq differ in $M_V$ only by $0^{\rm m}\!\!.09 \pm 0^{\rm m}\!\!.17$ \citep{Riess:2005}.

Na{\"i}vely, a shorter rise time would be associated with a larger photospheric expansion velocity $v_{\rm exp}$.  Near $t_{B{\rm max}}$, values of $v_{\rm exp}$ estimated from the Si II 6355-{\AA} minimum are provided for six of these SNe by \citet{Hachinger:2006}.  Of these six SNe, all but SN 2002bo belong to the low-expansion-velocity-gradient ({\sc lvg}) group described by \cite{Benetti:2005}; SN 2002bo, with a higher velocity gradient, belongs to the {\sc hvg} group.  As for the remaining two SNe, $v_{\rm exp}$ for SN 1998aq was observed by \citet{Vinko:1999}.  Its relative stability near $t_{B{\rm max}}$ suggests that SN 1998aq should also be classified as {\sc lvg}.  SN 2005cf was assigned to the {\sc lvg} group and its $v_{\rm exp}$ was measured by \citet{Garavini:2007}.  These eight values of $v_{\rm exp}$ exhibit a Pearson coefficient of correlation with fall time that is negligible, $R = -0.12^{+0.44}_{-0.39}$, but the correlation with rise time is more evident, $R = -0.39^{+0.43}_{-0.31}$.  The latter value differs from zero with the expected sign, but only at the one-standard-deviation level.               

\section{IMPLICATIONS FOR PRECISION COSMOLOGY} \label{impli}

\subsection {Rise Time Evolution} \label{evolu}

For use in measuring the expansion history of the Universe, the key aspect of any SN property is its possible evolution between high and low $z$.  As pointed out by \citet{Nugent:1998} and \citet{Riess:1999a}, evidence for substantial rise-time evolution could challenge the assumption that, after appropriate correction for light-curve widths, SN Ia absolute luminosities do not significantly evolve (for a recent discussion, see \citep{Howell:2007}).  The measurement of rise-time evolution was tackled by Con06, who used recent {\sc snls} data to achieve a large reduction in high-$z$ rise-time uncertainty.  Equally importantly for the high-$z$ {\it vs.}\ low-$z$ comparison, Con06 applied similar analysis tools, including SN light-curve templates, to both samples.  

Because the quality of high-$z$ SN photometry, despite recent advances, does not yet allow accurate definition of multicolor light curves for individual SNe without imposing external templates, the analysis reported here does not compete with Con06 in comparing high-$z$ to low-$z$ rise times.  The {\sc aquaa} contribution is only to suggest that intrinsic variance, especially at low $z$, should play a larger role in their error analysis.

\subsection {Correcting for Light-Curve Width} \label{width}

Apart from possible evolutionary effects, rise-time values influence measurements of distance moduli through their contribution to the light-curve widths for which SN absolute luminosities $M$ are corrected.  To explore this point, it is convenient to divide the $B$ light curve into pre-$t_{B{\rm max}}$ (pre-max) and post-max segments, using the rise time $t_{\rm r}$ and fall time $t_{\rm f}$, respectively, as measures of the pre-max and post-max segment widths.  The combination of $t_{\rm r}$ and $t_{\rm f}$ that effectively is used to correct $M_B$ depends on the choice of first parameter (\S \ref{decli}).  For example, $\Delta m_{15}$ depends only on $t_{\rm f}$, while, to a first approximation, stretch $s_B$ is proportional to a linear combination $t_{\rm r} + \gamma t_{\rm f}$ (\S \ref{chara}).  The value of $\gamma$ depends on the distribution of photometric errors over the time interval where the stretch fit is made; for the following discussion a representative value $\gamma = 2$ is adopted.  In \S\ref{intrin} a $\approx$ 1-day rms variation $\sigma^{\rm int}_{\rm rf}$ in the rise-time $-$ fall-time difference $t_{\rm rf}$ was reported, but the dependence on $t_{\rm rf}$ of $M_B$ is not yet known.

Choosing $t_{\rm f}$ and $t_{\rm rf}$ as the independent variables (\S\ref{corre}), one may rewrite 
\begin{equation}
s_B \propto t_{\rm f} + {1 \over \gamma\!+\!1} t_{\rm rf} 
\, . \label{eqsB}
\end{equation}
For correcting the absolute SN magnitude $M_B$, two simple possibilities are illustrative:  ({\it i}) $t_{\rm r}$ and $t_{\rm rf}$ are equally correlated with $M_B$; or ({\it ii}) $t_{\rm rf}$ is uncorrelated with $M_B$.  In case ({\it i}), correcting $M_B$ with any linear combination of rise time and fall time will yield essentially the same result.  In case ({\it ii}), results from correcting $M_B$ with stretch {\it vs.}\ correcting with fall time (which is tightly coupled to $\Delta m_{15}$) need not agree.  After proper calibration, these two choices will yield an rms intrinsic mutual difference in $M$: 
\begin{equation}
\sigma^{\rm int}_M \approx {\alpha \over \gamma\!+\!1} \, 
{\sigma^{\rm int}_{\rm rf} \over 15 \: {\rm days}}
  \approx 0^{\rm m}\!\!.033
\, , \label{eqsM}
\end{equation}
where $\alpha \approx 1.5$ is a typical stretch-correction coefficient.  Without more information, one cannot determine which of these two correction methods is better.  At present this choice is not of great practical consequence since, in quadrature with the $\approx 0^{\rm m}\!\!.18$ Hubble-line dispersion, $\sigma^{\rm int}_M$ is negligible.  

\subsection {Using a Template with a Biased Rise Time} \label{biase}

In analogy with equation (\ref{eqsM}), the systematic error made by correcting $M_B$ with a stretch that is measured using a template having a rise time that is too long by $\delta^{\rm sys}_{\rm rf} = +2$ days is
\begin{equation}
\delta^{\rm sys}_M \approx -{\alpha \over \gamma\!+\!1} \, 
{\delta^{\rm sys}_{\rm rf} \over 15 \: {\rm days}}
  \approx -0^{\rm m}\!\!.067
\, , \label{eqsys}
\end{equation}
which, representing $\approx 1/3$ of the magnitude of the dark energy signal at $z \approx 0.5$, is {\sl not} negligible.  Immunity to this source of systematic error often is sought by relying on the same template to fit high-$z$ and low-$z$ light curves, so that fitted values of high-$z$ and low-$z$ stretch would be biased in the same way.  

Unfortunately, if the high-$z$ and low-$z$ light curves are sampled differently, the cancellation in stretch bias is incomplete.  At high $z$, SNe often are discovered using a ``rolling'' strategy, in which the same sky area is imaged repeatedly with a rapid (few-day) cadence.  After a SN is discovered, its fluxes measured from stored pre-discovery images are added to the dataset, providing a uniformly sampled light curve.   At low $z$, a rolling strategy is observationally expensive due to the much larger sky area that must be imaged to monitor a comparable search volume; many nearby literature SNe were discovered and measured in other ways.  For example, of the 61 nearby SNe fitted by \citet{Wang:2006}, 22 (9) included no $B$ measurements before $t_{B{\rm max}}$ ($t_{B{\rm max}}+5$ days).

Consider a model experiment in which 100\% of high-$z$ and 60\% of low-$z$ light curves are sampled uniformly, while the remaining 40\% of low-$z$ light curves are sampled only after $t_{B{\rm max}}$.  From equation (\ref{eqsys}), using a template with a 2-day bias in rise time would introduce a systematic bias in the difference of high-$z$ and low-$z$ absolute SN magnitudes of $\approx 0^{\rm m}\!\!.027$.  Its order of magnitude is that of the largest single systematic error source considered in the {\sc snls} first-year paper \citep{Astier:2006}.

\section{PROSPECTS} \label{prosp}

In the short term, high priority is given to building a second, statistically independent version of the (presently unusual) distribution in Figs.~\ref{fig3} and \ref{fig4}.  Ideally, the SNe to be fitted for this purpose would share features with the best in the presently available sample:  well-observed, with $BV$ photometry available $\approx$2$^{\rm m}$ before $B_{\rm max}$ is reached, and sampled with $<4$-day cadence around $t_{B{\rm max}}$;  not too distant ($z < 0.02$), to minimize $K$-corrections;  not too dusty ($A_V < 1$), to minimize dereddening corrections; and not peculiar.  Full results of fits to shared photometry are shared with the provider.

It would be advantageous to have available a SN light-curve-fitting tool that (like {\sc aquaa}) is smooth and flexible enough to provide a statistically acceptable description of unusually well-observed photometric data, but that (unlike the present version of {\sc aquaa}) is also able gracefully and objectively to represent much coarser, sparser data.  A single tool could address these goals if its many fitted parameters were orthogonalized and mostly constrained by gentle priors derived from fits to the best-observed SNe.  

\acknowledgments

A substantial debt is owed Weidong Li, Andrea Pastorello, and Vallery Stanishev for providing photometric data prior to publication.  Essential to this study were the generous help and advice of many former and present members of the Supernova Cosmology Project, including Alex Conley, Gerson Goldhaber, Ariel Goobar, Don Groom, Alex Kim, Marek Kowalski, Peter Nugent, Reynald Pain, Saul Perlmutter, David Rubin, and Lifan Wang.  This work was supported by the Director, Office of Science, of the U.S.~Department of Energy under Contract No.~DE-AC02-05CH11231.

\appendix

\section{SUPERNOVA FIT DETAILS} \label{detai}

This Appendix collects miscellaneous facts and observations about photometric data for the SNe analyzed here, and about the {\sc aquaa} fits to those data.  This information is of possible interest to experts on these particular SNe.     

\subsection{SN 1994D}  \label{1994D}

The photometry used here is from \citet{Richmond:1995} and \citet{Patat:1996}.  In both papers, guidance on photometric uncertainties is provided, but point-by-point errors are not tabulated.  Therefore the iterated fitting procedure (\S\ref{apply}) used elsewhere for perturbing the errors was used here to set the error scale for each telescope and band.  Within a given telescope's dataset, errors also were varied in proportion to flux$^{-1/2}$ and in concert with the recorded seeing conditions.  It was checked that the fitted light-curve residuals, after normalization to these errors, exhibit no strong flux dependence. 

\subsection{SN 1998aq} \label{1998aq}

The earliest $B$ datum is only $\approx 1^{\rm m}\!\!.2$ below $B_{\rm max}$.  Fortunately, Rie99b transformed three unfiltered additional points to the $V$ band, including one that is $\approx 2^{\rm m}\!\!.8$ below $V_{\rm max}$.  (Not independently, Rie99b transformed these same points also to the $B$ band; they are not used here, as the $V$ filter more closely matches the unfiltered passband.)  Because two of the transformed $V$ points overlap well with two filtered $V$ measurements, {\sc aquaa} was able to allow the zero point of the unfiltered set to float freely; the fit is sensitive only to their relative magnitudes.  Nevertheless, the fitted zero point of the unfiltered data is in excellent agreement with that assigned by Rie99b.  

\subsection{SN 1990N}  \label{1990N}

As they did for SN 1998aq, Rie99b transformed eight unfiltered additional SN 1990N points to the $V$ band.  Of the  $0^{\rm m}\!\!.13$ error they assigned to those points, $0^{\rm m}\!\!.05$ was treated in the {\sc aquaa} fit as a point-to-point error.  Again, as for SN 1998aq, the transformed unfiltered points overlap well with two filtered $V$ measurements, so their zero point was left as a free parameter in the {\sc aquaa} fit.  Again it is in excellent agreement with the Rie99b value.

Gaps in observation of $\approx$8, $\approx$7, and $\approx$6 days occur between $t_{B{\rm max}} - 7$ and $t_{B{\rm max}} +22$ days.  The {\sc aquaa} fit became more robust when the stretch-corrected $B$ and $V$ spline knot phases were fixed to the values fitted for SN 2001el.  The quoted uncertainties include an allowance for the effect of applying this extra constraint.   

\subsection{SN 2001el} \label{2001el}

The host-galaxy-absorption part of the $R$-correction to SN 2001el used the values $A_V = 0^{\rm m}\!\!.54$ and $R_V = 2.15$ measured by \citet{Krisciunas:2007}, who compared SN 2001el to its less extinguished ``clone'' SN 2004S.  

\subsection{SN 2002bo} \label{2002bo}

The host-galaxy-absorption part of the $R$-correction to SN 2002bo used the value $R_V = 2.6$ (intermediate between Galactic and SN 2001el values) and a color excess ${\rm E}(\bv) = 0^{\rm m}\!\!.44$ derived from the {\sc aquaa} fit.  This yields $A_V = 1.14$, essentially the same value obtained by the optical photometry of \citet{Krisciunas:2004}.   

\subsection{SN 2003du} \label{2003du}

The earliest point from {\it HCT} \citep{Anupama:2005}, $\approx 1^{\rm m}\!\!.75$ below $B_{\rm max}$, lies 7.6 quoted standard deviations from a smooth curve drawn through nearby points from {\it HCT} and from the {\it NOT} and {\it Asiago} 1.8m telescopes \citep{Stanishev:2007}; the other 42 points lie within 2.3$\sigma$ of the fitted curve.  Therefore its uncertainty was increased to $\pm 0^{\rm m}\!\!.15$ (see Fig. \ref{fig1}).  In this isolated case, the assumption that the fitted $B$ light curve is smooth led to the use of information from later points to reduce the relative weight of an earlier point.      

\subsection{SN 2004eo} \label{2004eo}

The photometry is from \citet{Pastorello:2007a} and from the preliminary $B$ and $V$ points shown in Fig.~5 of \citet{Hamuy:2006}.  Independent fits to each dataset were performed; all quoted parameters are unweighted averages of the two fit values.  The quoted uncertainties include allowances for possible systematic differences between the two components.

The $K$- and $S$-corrections to the data of \citet{Pastorello:2007a} are taken from their paper, while those to the data of \citet{Hamuy:2006} are discussed in \S\ref{sampl}.  To the latter points, which are those shown in Fig.\ \ref{fig1}, uncertainties of $0^{\rm m}\!\!.015$ were assigned at peak; elsewhere they grew in proportion to flux$^{-1/2}$.   Again it was checked that the normalized light-curve residuals exhibit no strong flux dependence.

\subsection{SN 2005cf} \label{2005cf}

The $B$ magnitude at JD 2,453,542.53 from \citet{Pastorello:2007b} lies 7.1 quoted standard deviations from the fitted curve; the other 38 points lie within 1.8$\sigma$ of the curve.  Therefore its uncertainty was increased to $\pm 0^{\rm m}\!\!.15$ (see Fig. \ref{fig1}).    

\section{COMPARISON OF PUBLISHED RISE TIMES OF INDIVIDUAL SN{\lowercase{e}} TO AQUAA VALUES} \label{compa}


\input{tab2.tex}


\noindent
Table \ref{table2} collects published rise times for five SNe studied in this paper.  Also tabulated are the times of maximum $B$ flux from the published rise time analyses.  Excluding the measurements of SN 1990N by \citet{Leibundgut:1991} and of SN 1994D by \citet{Vacca:1996}, in which a unique rise-time function was used, on average the published rise times exceed those of {\sc aquaa} by $1.28 \pm 0.33$ days.  Of this excess, an average of $0.44 \pm 0.15$ days may be attributed to the later times of maximum $B$ flux used in the published analyses; this point was discussed in \S{\ref{riset}}.  There the remaining $\approx 0.8$-day excess was ascribed to two properties of the piecewise analysis method of Rie99b used in those publications:  at $t_{\rm join}\approx 10$ days before $t_{B{\rm max}}$, the two fitted pieces of the light curve are not required to join; and, for $t < t_{\rm join}\,$, $\bv$ color is assumed not to vary.  Here the implications of these two properties are illustrated.

Consider the plot of $B$-band root flux {\it vs.}\ time shown for SN 2004eo in Fig.\ \ref{fig1}; suppose that a linear fit is made to early points.  If, as for {\sc aquaa}, fluxes only within the first 40\% of the rise time are assumed to rise parabolically, only the earliest three points belong in this fit.  On the other hand, if all five points up to $t_{B{\rm max}} - 8.9$ days are included, $\chi^2$ is still acceptable (4.7/3 dof), but the fitted time of explosion is earlier by 0.4 days.  If they are fitted in isolation, all five points are not badly represented by a straight line; however, if (as by {\sc aquaa}) they are fitted as part of a full light curve with continuous 0$^{\rm th}$ through 3$^{\rm rd}$ derivatives, the portion of the best-fitted curve that approximates a straight line is shorter, as is the fitted rise time itself.

Secondly, consider the $V$-band plot for SN 1990N in Fig.\ \ref{fig1}, and suppose that a linear fit is made to the points with $\tau < -10$ days.  However, following equation (\ref{eqfX}) and the discussion in \S\ref{appro}, if the $B$-band rise is parabolic and if the best-fit initial rate of change of $\bv$ color is negative, as it is for SN 1990N, the curve that instead should be fitted to these points is slightly concave downward, as shown in Fig.\ \ref{fig1}.  Both fitted values of $\chi^2$ are equally acceptable, but the straight line yields a fitted time of explosion that is 0.3 days earlier than that of the curve.   

{}

\end{document}

%% file: tab1.tex
\begin{deluxetable*}{lccccc}  
\tablecolumns{6}
\tableheadfrac{}
\tablecaption{Results of AQUAA light-curve fits.\label{table1}}
\tabletypesize{\footnotesize}  
\tablehead{\colhead{SN}
&\colhead{\begin{tabular}{c} $t_{B{\rm max}}$ \\ (${\rm JD} - 2$,440,000) \end{tabular}}
&\colhead{\begin{tabular}{c} $\Delta m_{15}$ \\ (mag) \end{tabular}}
&\colhead{\begin{tabular}{c} Rise time \\ (rest-frame \\ days) \end{tabular}}
&\colhead{\begin{tabular}{c} Rise time \\ $-$ fall time \\ (rest-frame \\ days) \end{tabular}}
&\colhead{\begin{tabular}{c} Photometry \\ references \end{tabular}}
}
\startdata
SN 1990N           & $\;\,8082.46 \pm 0.35$ & $0.990 \pm 0.034$ & $20.01 \pm 0.46$ & $4.04 \pm 0.74$ & $ (1,2)     $\\
SN 1994D           & $\;\,9432.29 \pm 0.14$ & $1.344 \pm 0.021$	& $15.39 \pm 0.47$ & $2.04 \pm 0.53$ & $ (3,4)     $\\
SN 1998aq	      & $10931.04 \pm 0.14$ & $1.042 \pm 0.021$	& $17.52 \pm 0.58$ & $2.03 \pm 0.64$ & $ (5,2)     $\\
SN 2001el	      & $12182.19 \pm 0.17$ & $1.168 \pm 0.021$	& $18.00 \pm 0.56$ & $3.57 \pm 0.63$ & $ (6,7)     $\\
SN 2002bo	      & $12356.39 \pm 0.20$ & $1.162 \pm 0.031$	& $16.05 \pm 0.37$ & $1.57 \pm 0.53$ & $ (8,9,10)  $\\
SN 2003du	      & $12765.95 \pm 0.14$ & $0.982 \pm 0.021$	& $17.71 \pm 0.35$ & $1.64 \pm 0.44$&$(11,12,13,14)$\\
SN 2004eo	      & $13278.65 \pm 0.14$ & $1.403 \pm 0.037$	& $16.64 \pm 0.44$ & $3.86 \pm 0.53$ & $(15,16)    $\\
SN 2005cf	      & $13533.78 \pm 0.14$ & $1.068 \pm 0.021$	& $16.62 \pm 0.25$ & $1.35 \pm 0.36$ & $(17)       $\\
Fiducial (slow rising\tablenotemark{a})& \nodata & $1.1$        & $18.81 \pm 0.36$ & $3.81 \pm 0.36$ & \nodata      \\
Fiducial (fast rising\tablenotemark{b})& \nodata & $1.1$        & $16.64 \pm 0.21$ & $1.64 \pm 0.21$ & \nodata      \\
Fiducial (all\tablenotemark{c})	       & \nodata & $1.1$	& $17.44 \pm 0.39$ & $2.44 \pm 0.39$ & \nodata      \\
\enddata
\tablenotetext{a}{Average of SN 1990N, SN 2001el, and SN 2004eo, corrected to $\Delta m_{15} = 1^{\rm m}\!\!.1$.}
\tablenotetext{b}{Average of all but SN 1990N, SN 2001el, and SN 2004eo, corrected to $\Delta m_{15} = 1^{\rm m}\!\!.1$.}
\tablenotetext{c}{Average of all SNe, corrected to $\Delta m_{15} = 1^{\rm m}\!\!.1$, and with errors augmented to account for intrinsic variance.}
\tablerefs{(1) \citet{Lira:1998}; (2) \citet{Riess:1999b}; (3) \citet{Richmond:1995}; (4) \citet{Patat:1996}; (5) \citet{Riess:2005}; (6) \citet{Krisciunas:2003}; (7) \citet{Krisciunas:2007}; (8) \citet{Benetti:2004}; (9) \citet{Krisciunas:2004}; (10) \citet{Szabo:2003}; (11) \citet{Leonard:2005}; (12) \citet{Anupama:2005}; (13) \citet{Stanishev:2007}; (14) Li, W.D.\ \etal\ (2007), in preparation; (15) \citet{Hamuy:2006}; (16) \citet{Pastorello:2007a}; (17) \citet{Pastorello:2007b}.}
\end{deluxetable*}  

%% file: tab2.tex
\begin{deluxetable*}{llclcc}  
\tablecolumns{6}
\tableheadfrac{}
\tablecaption{Comparison of published $B$-band times of maximum flux and rise times to {\sc aquaa} values for individual SNe.\label{table2}}
\tabletypesize{\footnotesize}  
\tablehead{\colhead{SN}
&\colhead{\begin{tabular}{c} Published\tablenotemark{a} \\ $t_{B{\rm max}}$ \\ (JD$-$244000) \end{tabular}}
&\colhead{\begin{tabular}{c} Published\tablenotemark{a} \\ $-$ {\sc aquaa} \\ $t_{B{\rm max}}$ \\ (days) \end{tabular}}
&\colhead{\begin{tabular}{c} Published \\ $B$ rise time \\ (rest-frame \\ days) \end{tabular}}
&\colhead{\begin{tabular}{c} Published \\ $-$ {\sc aquaa} \\ $B$ rise time \\ (rest-frame \\ days) \end{tabular}}
&\colhead{\begin{tabular}{c} Rise time \\ reference \end{tabular}}
}
\startdata
SN 1990N  & $\;\;\;\; 8083.0 \pm 1.0$          & $ 0.5 $ & $\;\;\; 18.5 \pm 1.5$\tablenotemark{b} & $ -1.5\;\;\,$ & $ (1)     $ \\
SN 1990N  & $\;\;\;\; 8083.4$\tablenotemark{c} & $ 0.9 $ & $\;\;\; 21.4 \pm 0.3$                  & $ 1.4 $       & $ (2)     $ \\
SN 1994D  & $\;\;\;\; 9432.9 \pm 0.2$          & $ 0.6 $ & $\;\;\; 17.6 \pm 0.5$\tablenotemark{d} & $ 2.2 $       & $ (3)     $ \\
SN 1994D  & $\;\;\;\; 9432.7$\tablenotemark{e} & $ 0.4 $ & $\;\;\; 15.5 \pm 0.5   $               & $ 0.1 $       & $ (2)     $ \\
SN 2002bo & $\;\; 12356.5 \pm 0.5$             & $ 0.1 $ & $\;\;\; 17.9 \pm 0.5   $               & $ 1.8 $       & $ (4)     $ \\
SN 2004eo & $\;\; 13279.2 \pm 0.5$             & $ 0.5 $ & $\;\;\; 17.7 \pm 0.6   $               & $ 1.1 $       & $ (5)     $ \\
SN 2005cf & $\;\; 13534.0 \pm 0.3$             & $ 0.2 $ & $\;\;\; 18.6 \pm 0.4   $               & $ 2.0 $       & $ (6)     $ \\
Average\tablenotemark{f} & $\;\;\;$ \nodata  & $0.44 \pm 0.15$ & $\;\;\;$ \nodata  & $1.28 \pm 0.33$ & \nodata  \\
RMS\tablenotemark{f}     & $\;\;\;$ \nodata  & $0.33$          & $\;\;\;$ \nodata  & $0.75$          & \nodata  \\
\enddata
\tablenotetext{a}{Value quoted by or deduced from information in rise time reference.}
\tablenotetext{b}{A range of 17-20 days is quoted in reference (1).}
\tablenotetext{c}{Estimate deduced by identifying the last two entries for SN 1990N in Table 4 of reference (2) with the entries for ${\rm JD}-2440000 = 8071.5$ and 8072.5 in Table 5 of \citet{Lira:1998}.}
\tablenotetext{d}{In a more recent analysis using methods similar to those of reference (3), \citet{Contardo:2000} obtained a best fit bolometric rise time for SN 1994D that was 1.4 days shorter than that of reference (3).  If the $B$ rise time were similarly reduced, it would fall within 1 day of the {\sc aquaa} value.}
\tablenotetext{e}{Estimate deduced by identifying the first and third entry for SN 1994D in Table 4 of reference (2) with the first two entries in Table 5 of \citet{Richmond:1995}.}
\tablenotetext{f}{Excludes references (1) and (3), in which a unique rise time function was used.}
\tablerefs{(1) \citet{Leibundgut:1991}; (2) \citet{Riess:1999b}; (3) \citet{Vacca:1996}; (4) \citet{Benetti:2004}; (5) \citet{Pastorello:2007a};  (6) \citet{Pastorello:2007b}.}
\end{deluxetable*}  